\documentclass{aastex63}

\usepackage{graphicx,subfigure,color,amsmath,pifont,multirow,threeparttable}
\usepackage{colortbl}
\usepackage{float}
\floatstyle{plain}

\definecolor{mycyan}{rgb}{1.00,0.91,0.86}

\submitjournal{ApJ}

\shortauthors{Zhang et al.} 

\usepackage{hyperref} 
\usepackage{cleveref} %
\usepackage{bm}

\begin{document}
\title{Photospheric Emission of an Accelerating/a Decelerating Relativistic Jet}
\correspondingauthor{Da-Bin Lin}     
\email{lindabin@gxu.edu.cn}   
\author{Hao-Qiang Zhang}
\affil{Guangxi Key Laboratory for Relativistic Astrophysics, School of Physical Science and Technology, Guangxi University, \\Nanning 530004, China}
\author{Da-Bin Lin}
\affil{Guangxi Key Laboratory for Relativistic Astrophysics, School of Physical Science and Technology, Guangxi University, \\Nanning 530004, China}
\author{Guo-Yu Li}
\affil{Guangxi Key Laboratory for Relativistic Astrophysics, School of Physical Science and Technology, Guangxi University, \\Nanning 530004, China}
\author{Zhi-Lin Chen}
\affil{Guangxi Key Laboratory for Relativistic Astrophysics, School of Physical Science and Technology, Guangxi University, \\Nanning 530004, China}

\begin{abstract}
In the gamma-ray bursts (GRBs), Compton scattering near the photosphere plays a crucial role in shaping the observed jet emission.
Previous studies have considered the propagation of photons in a coasting relativistic jet.
While previous studies have primarily focused on coasting relativistic jets, 
the behavior of photons in accelerating or decelerating jets has remained relatively unexplored. 
In this work, we investigate the photospheric emission of an accelerating/a decelerating relativistic jet, with the Lorentz factor $\Gamma$ described as $\Gamma\propto r^{\delta}$, where $r$ is the distance relative to the central engine of the GRB and $\delta$ is the index.
We derive an analytical solution showing that, when the optical depth is larger than unity, the comoving photon energy evolves as $\epsilon'(r) \propto r^{-(2+\delta)/3}$, which differs significantly from the coasting jet scenario.
Using Monte Carlo Radiation Transfer (MCRaT) simulations, we numerically track the energy evolution of photons, which confirms our analytical results for $\tau \gg 1$ but reveals a deviation from the previously assumed power-law form near $\tau \approx 1$.
Furthermore, we calculate the resulting radiation spectra and find that, compared to coasting jets, the spectrum significantly broadens in decelerating jets ($\rm\delta = -1/3$) while narrowing in accelerating jets ($\rm\delta = 1/3$).
We find that this spectral broadening is primarily driven by the weakening of relativistic beaming in a decelerating jet.

\end{abstract}
\keywords{Gamma-ray bursts; Relativistic jets; Thermal–radiative transfer–scattering }

\section{Introduction}\label{Introduction}
Relativistic jets exist in various types of astronomical objects, such as gamma-ray bursts (GRBs; \citealp{2006MNRAS.369.1287G, 1986ApJ...308L..47G}),  microquasars (\citealp{1994Natur.371...46M, 1995Natur.375..464H}) and active galactic nuclei (AGNs; \citealp{1985ApJ...295..358L, 2006MNRAS.369.1287G}).
In such an astrophysical environment, the density of the internal plasma may be very high, and thus the optical depth of Thomson scattering of photons is considerable.
When the Thomson scattering optical depth exceeds unity, photons are effectively trapped within the plasma.
As the relativistic jet propagates outward, the plasma density gradually decreases, causing the optical depth to decrease, and eventually, these trapped photons finally escape at a position where the optical depth is roughly unity (i.e., the photospheric radius $r_\mathrm{ph}$).
Consequently, the emergent radiation spectrum typically exhibits a blackbody or quasi-blackbody profile.

Interestingly, based on analysis of the observed spectral shapes,
quasi-thermal components were identified in some GRBs (\citealp{2004ApJ...614..827R, 2005ApJ...625L..95R, 2009ApJ...706L.138A, 2010ApJ...709L.172R, 2011ApJ...727L..33G, 2012ApJ...757L..31A, 2013MNRAS.432.3237G, 2015ApJ...800L..34L, 2022ApJ...934L..22D}), 
in particular, the thermal components of  GRB 090902B \cite{2009ApJ...706L.138A} and GRB 220426A \cite{2022ApJ...934L..22D} were particularly significant.
Therefore, the photospheric model was proposed to explain the thermal component of these GRBs (e.g., \citealp{1991ApJ...369..175A, 1994MNRAS.270..480T, 2000ApJ...530..292M, 2005ApJ...628..847R, 2011ApJ...731...80N, 2012ApJ...755L...6F, 2013ApJ...765..103L, 2013ApJ...772...11R, 2015ApJ...801..103G, 2015ApJ...802..134B, 2015ApJ...813..127P, 2017MNRAS.472.1897R, 2018MNRAS.475.1708A, 2018ApJ...866...13H, 2018ApJ...860...72M, 2019ApJ...882...26M, 2019ApJS..242...16L, 2019ApJS..245....7L, 2019ApJ...884..109L, 2020ApJ...893..128A, 2020ApJ...899..111W, 2021ApJ...922..237W, 2022MNRAS.509.6047M}).
In the photospheric model, the radial evolution of the photon energy significantly influences the resulting radiation spectrum. 
It is known that when the Thomson scattering optical depth is sufficiently high, electrons and photons in the outflow achieve thermal equilibrium through multiple scatterings. 
As a result, previous studies have typically assumed that the photon and electron energy are tightly coupled (e.g., \citealt{2020ApJ...899..111W,  2022MNRAS.509.6047M}).
However, \cite{2008ApJ...682..463P} demonstrated that the photon energy is also modulated by Compton scattering. Specifically, a misalignment between the photon scattering direction and the electron velocity vectors induces significant photon cooling. 
This cooling mechanism becomes dominant when the electrons cool faster, thereby substantially altering the spectral shape.

The central engines of GRBs are widely thought to power relativistic jets, which can undergo acceleration in radiation pressure-dominated (thermal photon) or Poynting-flux-dominated regimes (e.g., \citealt{2011MNRAS.413.2031M,2013ApJ...765..125L}). 
Furthermore, these jets may experience deceleration during propagation through the circum-burst medium, as discussed by \citet{granot2012interaction}. 
Given these dynamics, a dedicated study of photon cooling mechanisms in accelerating or decelerating relativistic jets is imperative to fully understand their emission properties.
However, \cite{2008ApJ...682..463P} explored the scenario of stable outflows but did not delve into the cooling behavior of photons in relativistic jets subject to acceleration or deceleration. 
This paper studies the cooling behavior of photons due to Compton scattering in the acceleration and deceleration scenarios of relativistic jets. 
Meanwhile, as the relativistic jet continues to expand, its optical depth gradually decreases,
primarily due to the declining density of the outflow \cite{2020ApJ...899..111W}. 
Upon reaching the photospheric radius, the scattering probability between photons and electrons becomes extremely low, rendering them unable to exchange energy sufficiently. 
This ultimately results in thermal decoupling, at which point the photon energy approaches a stable state (as assumed in Equation~(\ref{eq:temperature_evolution}) of this paper).
Additionally, we utilize the Monte Carlo Radiation Transfer (MCRaT) to conduct numerical simulations of these scenarios and compare the results with analytical estimations. 
The analytically derived predictions of photon energy evolution with radius and spectral broadening due to jet deceleration are both corroborated by our numerical simulations.

The paper is organized as follows: Section~\ref{sec2} describes analytical estimations of photon energy evolution with radius in the accelerating and decelerating jets, 
followed by numerical simulations using MCRaT. 
Section~\ref{sec3} presents radiation spectra of photospheric emission from accelerated and decelerated jets, respectively, obtained through analytical estimation and MCRaT's simulations, including a comparison between the two results. 
Section~\ref{discussion and summary} summarizes key findings and discusses implications.

\section{Photon's energy evolution in relativistic jets}\label{sec2}
This paper focuses on photospheric emission from relativistic jets that are either accelerating or decelerating. 
To begin with, the evolution of the Lorentz factor during the jet's expansion needs to be defined. 
Two primary mechanisms can accelerate a gamma-ray burst jet: photon radiation pressure ($\Gamma \propto r^{\delta}$, $\delta = 1/3$) and magnetic pressure acceleration ($\Gamma \propto r$, $\delta = 1$) \cite{2009MNRAS.394.1182K, 2011MNRAS.411.1323G}. 
Here, $\Gamma$ represents the Lorentz factor of the jet, $r$ is the distance from the central engine, and $\delta$ is the power-law index. 
In contrast, if the relativistic jet is decelerated by the circum-burst medium, a negative power-law index is expected ($\delta < 0$). 
In this work, we adopt $\delta = -1/3$ for the deceleration phase.

\subsection{Analytical Estimation on the evolution of Photon's Energy in Relativistic Jets.}
Following \cite{2008ApJ...682..463P}, we assume that the comoving electron temperature is negligible ($T'_{\rm el} = 0$). 
Hereafter, primed (unprimed) quantities denote measurements in the jet comoving frame (observer frame). 
The geometry defined in Figure 2 in \cite{2008ApJ...682..463P} forms the basis of our setup.
We adopt a cylindrical coordinate system centered on the jet axis, with the line of sight aligned along the positive $z$-axis.

Consider a photon propagating toward the observer that undergoes its $(i+1)$th scatter at position $r_{i+1}$ after the $i$th scatter at $r_i$, where $r_{i}=(r_{\rm min}^2+z_{i}^2)^{1/2}$ and $r_{i+1}=(r_{\rm min}^2+z_{i+1}^2)^{1/2}$
are the distances from the $i$th and the ($i+1$)th scattering sites to the center, respectively,
$z_{i}$ and $z_{i+1}$ are the projection distances of $r_{i}$ and $r_{i+1}$ on the $z-$axis, respectively.
The bulk Lorentz factor of the jet at $r_{i}$ and $r_{i+1}$ are $\Gamma_{i}$ and $\Gamma_{i+1}$, respectively.
After the $i$th scatter, the photon energy is assumed to be $\epsilon'_{i}$
and the photon direction is at the angle $\theta_{\gamma}$ with respect to the $i$th scattered electron velocity vector (in the observer frame).
Then, we obtain 
\begin{equation}\label{EQ:photon energy A}
\epsilon_i = \frac{\epsilon'_{i}}
{\Gamma_{i}\left[1-\beta_{i}\cos\theta_{\gamma}\right]}.
\end{equation}
Assuming that the electron participating in the $(i+1)$th scatter
has an angle $\theta_{\rm el}$ with the velocity vector of the electron participating in the $i$th scatter,
one can have 
\begin{equation}\label{EQ:photon energy B}
\epsilon'_{i+1} = \epsilon_i
\Gamma_{i+1}\left[1-\beta_{i+1}\cos(\theta_{\gamma}-\theta_{\rm el})\right].
\end{equation}
Substituting Equation~(\ref{EQ:photon energy A}) into Equation~(\ref{EQ:photon energy B}) yields:
\begin{equation}\label{EQ:photon energy C}
\epsilon'_{i+1} = \epsilon'_{i}(\frac{\Gamma_{i+1}}{\Gamma_{i}})\frac{1}{1-\beta_{i}\cos\theta_{\gamma}}
\left[1-\beta_{i+1}\cos(\theta_{\gamma}-\theta_{\rm el})\right].
\end{equation}
With 
\begin{multline}
\left[1-\beta_{i+1}\cos(\theta_{\gamma}-\theta_{\rm el})\right] = \Gamma_{i+1}^2\left[1-\beta_{i+1}^2\cos\theta_{\rm el}\right]
\left[1-\beta_{i+1}\cos\theta_{\gamma}\right]\\
-\Gamma_{i+1}^2\beta_{i+1}\left[\cos\theta_{\rm el}-1\right]
\left[\cos\theta_{\gamma}-\beta_{i+1}\right]
-\beta_{i+1}\sin\theta_{\rm el}\sin\theta_{\gamma},
\end{multline}
Equation~(\ref{EQ:photon energy C}) becomes
\begin{equation}\label{EQ:photon energy D}
\epsilon'_{i+1} =
\epsilon'_{i}(\frac{\Gamma_{i+1}}{\Gamma_{i}})\left[1+\frac{\beta_{i}-\beta_{i+1}}
{1-\beta_{i}\cos\theta_{\gamma}}\cos\theta_{\gamma}
+\frac{\beta_{i+1}(1-\cos\theta_{\rm el})}{1-\beta_{i}\cos\theta_{\gamma}}\cos\theta_{\gamma}
-\beta_{i+1}\sin\theta_{\rm el}
\frac{\sin\theta_{\gamma}}{1-\beta_{i}\cos\theta_{\gamma}}\right].
\end{equation}
According to the geometric relationship, the relation between $\theta_{\rm el}$ and $\theta_{\gamma}$ is
\begin{equation}\label{EQ:photon radii A}
\tan\theta_{\rm el} = \frac{(\Delta r_{i})\sin\theta_{\gamma}}
{r_{i}+(\Delta r_{i})\cos\theta_{\gamma}},
\end{equation}
where $\Delta r_{i}=z_{i+1}-z_{i}$ is the distance between the $i$th scattering site and the ($i+1$)th scattering site.
Note that below the photosphere, the average distance between two consecutive scatterings of the photon is very small ${\Delta r_{i}}/{r_{i}}\ll1$.
Thus, we can have an approximate $\tan\theta_{\rm el} \approx {(\Delta r_{i}}/{r_{i}})\sin\theta_{\gamma}$ and $\cos\theta_{\rm el} \approx 1- \theta_{\rm el}^2/2 \approx 1-({\Delta r_{i}}/{r_{i}})^2\sin^2\theta_{\gamma}/2$.
Since $\Gamma_{i+1}/\Gamma_{i}=({r_{i+1}/r_{i}})^{\delta}$, Equation~(\ref{EQ:photon energy D}) can be approximated as
\begin{equation}\label{EQ:photon energy E}
\epsilon'_{i+1} \approx \epsilon'_{i}
\left[1+\delta\frac{\Delta r_{i}}{r_{i}}\cos\theta_{\gamma}-
\frac{2\delta}{2\Gamma_{i}^2(1-\beta_{i}\cos\theta_{\gamma})}
\frac{\Delta r_{i}}{r_{i}}\cos^2\theta_{\gamma}
-\beta_{i+1}\frac{\Delta r_{i}}{r_{i}}
\frac{\sin^2\theta_{\gamma}}{1-\beta_{i}\cos\theta_{\gamma}}\right].
\end{equation}
As shown in Equation~(\ref{EQ:photon energy E}), the photon energy loss in each scatter depends crucially on the value of $\Delta r_{i}/r_{i}$.
The optical depth measured along a ray traveling in the positive z-axis and reaching the observer is given by
\begin{equation}
\tau=\int_{z_{i}}^{\infty}\frac{\dot{M}}{4\pi m_p \beta_{i} c \Gamma_{i} r^2_{i}}\sigma_{\rm T}\Gamma_{i}(1-\beta_{i}\cos\theta_{\gamma})dz,
\end{equation}
where $\dot{M}$ is mass-loss rate of the ejection of a spherically symmetric plasma outflow, $m_p$ is the rest mass of the proton and $\sigma_{\rm T}$ is the Thompson scattering cross section.
Considering a small distance $\Delta z$, the optical depth near this position can be expressed as
\begin{equation}
\begin{split}
\Delta\tau_{i}&=\Delta z\frac{\dot{M}}{4\pi m_p \beta_{i} c \Gamma_{i} r^2_{i}}\sigma_{\rm T}\Gamma_{i}(1-\beta_{i}\cos\theta_{\gamma})\\
&=\Delta r_{i}\frac{R_{\rm d}}{\pi r^2_{i}\beta_{i}}(1-\beta_{i}\cos\theta_{\gamma}),
\end{split}
\end{equation}
where $R_{\rm d}=\dot{M}\sigma_{\rm T}/(4 m_pc)$.
Since, on the average $\langle \tau_{i} \rangle=\langle \Delta r_{i}
R_{\rm d}(1-\beta\cos\theta_{\gamma})/(\pi r^2_{i}\beta_{i}) \rangle=1$, one obtains
\begin{equation}\label{EQ:photon energy F}
\begin{split}
\langle \Delta r_{i} \rangle &= \left[\frac{R_{\rm d}}{\pi r^2_{i}\beta_{i}}(1-\beta_{i}\cos\theta_{\gamma})\right]^{-1}\\
&=\left[\frac{R_{\rm d}}{\pi r^2_{i}\beta_{i}}\mu_i\right]^{-1},
\end{split}
\end{equation}
where $\mu_i = 1-\beta_{i}\cos\theta_{\gamma}$.
With Equation~(\ref{EQ:photon energy F}),
Equation~(\ref{EQ:photon energy E}) can be
reduced to
\begin{equation}\label{EQ:photon energy G}
\begin{split}
\epsilon'_{i+1} &\approx \epsilon'_{i}\{
1+\frac{\Delta r_{i}}{r_{i}}\left[\delta \frac{1-\mu_i}{\beta_{i}}
-\frac{\delta}{\Gamma^2_{i}} \mu_i^{-1}(\frac{1-\mu_i}{\beta_{i}})^2
-\beta_{i+1}\mu_i^{-1} +\beta_{i+1}\mu_i^{-1}(\frac{1-\mu_i}{\beta_{i}})^2 \right]\}\\
&\approx \epsilon'_{i}\{1+
(\frac{R_{\rm d}}{\pi r_{i}}\mu_i)^{-1}
\left[\delta -2\frac{\beta_{i+1}}{\beta_{i}}-\mu_i^{-1}\frac{1}{\Gamma^2_{i}}
\frac{1}{\beta_{i}}(\delta-\beta_{i+1}) +\frac{2\delta}{\Gamma^2_{i}}\frac{1}{\beta_{i}} -\mu_i\delta -\frac{\delta}{\Gamma^2_{i}}\frac{1}{\beta_{i}}\mu_i +\frac{\beta_{i+1}}{\beta_{i}}\mu_i
 \right]\}.
\end{split}
\end{equation}
Note that below the photosphere,  the photon scattering angle
$\theta_{\gamma} \sim \Gamma^{-1} \ll1$ in the observer frame can be find $\mu_i=O(\Gamma^{-2})$.
The small scatter angle implies that the radii of the consequent scatterings can be approximated as $r_{i+1}-r_{i} \simeq z_{\rm max}-z_{\rm min}$.
In addition, since ${\Delta r_{i}}/{r_{i}}\ll1$, one can have $\beta_{i+1}/\beta_{i} \simeq 1$.
Using these approximations in Equation~(\ref{EQ:photon energy G}), it can be approximated as
\begin{equation}\label{EQ:photon energy H}
\epsilon'_{i+1}
\approx \epsilon'_{i}\{1+
(\frac{R_{\rm d}}{\pi r_{i}}\mu_i)^{-1}
\left[\delta -2\frac{\beta_{i+1}}{\beta_{i}}-\mu_i^{-1}\frac{1}{\Gamma^2_{i}}
\frac{1}{\beta_{i}}(\delta-\beta_{i+1})
 \right]\}.
\end{equation}

So far, we have presented the situation in which a photon undergoes scattering.
In reality, there are a group of photons scattered by electrons.
In the spirit of \cite{2008ApJ...682..463P},
the average energy loss of this group of photon scatterings can be obtained
\begin{equation}\label{EQ:photon energy I}
\begin{split}
\epsilon'_{i+1} &\approx
\epsilon'_{i}\left[1+(\frac{R_{\rm d}}{\pi r_{i}})^{-1}\Gamma^2_{i}
(-\frac{1}{3}\delta-2+\frac{4}{3})\right]\\
&\approx
\epsilon'_{i}\left[ 1+(\frac{R_{\rm d}}{\pi r_{i}})^{-1}\Gamma^2_{i}
(-\frac{2+\delta}{3})\right]\\
&\approx
\epsilon'_{i}\left[1+\langle \omega \rangle\right],
\end{split}
\end{equation}
where $\langle \mu_i \rangle^{-1}=\Gamma_i^2$, $\langle \mu_i \rangle^{-1} \simeq 4\Gamma_i^4/3$, and $\beta_{i+1}/\beta_{i} \simeq 1$ are adopted,
and $\langle \omega \rangle= [R_{\rm d}/(\pi r_{i})]^{-1}\times$
$\Gamma^2_{i}[-(2+\delta)/{3}]$. Then, one can have
\begin{equation}\label{EQ:photon energy}
\epsilon'_{n} \approx \epsilon'_{1} e^{-n r_{1}\langle \omega \rangle}
\end{equation}
Based on the assumption further in Equations~(\ref{EQ:photon energy}) and Equations~(\ref{EQ:photon radii A}),
the group of photons average comoving energy and radius after $n$ scattering events can be approximated as
\begin{equation}\label{EQ:photon radii B}
r_{n} \approx r_{1} e^{n r_{1}\langle \alpha \rangle},
\end{equation}
where $\langle \alpha \rangle = [R_{\rm d}/(\pi r_{i})]^{-1}\Gamma^2_{i}$.
Finally, using the results of Equations~(\ref{EQ:photon radii B}),
one can find that the comoving energy of photons decreases with radius as
\begin{equation}\label{EQ:photon energy J}
\epsilon'_{\rm r} \propto r^{\langle \omega \rangle/\langle \alpha \rangle}=r^{-(\frac{2+\delta}{3})}.
\end{equation}
Based on Equations~(\ref{EQ:photon energy J}), the evolution of the photon energy with radius in the observer frame can be expressed as 
\begin{equation}\label{EQ:photon energy K}
\epsilon_{\rm r} \propto r^{[-(\frac{2+\delta}{3})+\delta]}.
\end{equation}
It should be emphasized that the conclusion of Equation~(\ref{EQ:photon energy J}) is obtained based on the assumption of ${\Delta r_{i}}/{r_{i}}\ll1$,
that is, the conclusion is only valid below the photosphere.

\subsection{Numerical results of the photon energy evolution in relativistic jets}\label{sec2.2}
We use the MCRaT to simulate the evolution of photon energy with radius in the accelerated and decelerated relativistic jets.
The MCRaT is a radiation transfer tool based on hydrodynamic simulations  (\citealp{2016ApJ...829...76L, 2018ApJ...853....8P, 2018ApJ...869..103P, 2021ApJ...922..257P, 2022ApJ...926..104P}).
The jet is configured with the following parameters:
the dimensionless entropy of the jet $\eta=400$;
the photons and electrons are injected at the initial radius $r_{\rm 0}= 10^{8}\, {\rm cm}$, 
and simulation ends at the radius $ r_{\mathrm{end}}(\delta=0) = r_{\mathrm{end}}(\delta=1/3) = 2\times10^{14}\, {\rm cm}$  for the coasting and acceleration cases, and at the radius $ r_{\mathrm{end}}(\delta=-1/3) = 2\times10^{16}\, {\rm cm}$ for the deceleration cases; 
the initial Lorentz factor $\Gamma_{0}=\eta$,
and the evolution of the Lorentz factor of the jet with radius is described by the following.
\begin{equation}\label{Eq:Gamma_evolution}
\Gamma=\left\{\begin{array}{ll}
\frac{r}{r_{\rm 0}}, & r < r_{\rm s},\\
\\
\frac{r_{\rm s}}{r_{\rm 0}} (\frac{r}{r_{\rm s}})^{\delta}, & r \geq r_{\rm s},
\end{array}\right.
\end{equation}
where $r_{\rm s}=\eta r_{\rm 0}$ is the saturation radius;
the half-open angle of the jet is taken as $\theta_{\rm j}=10/\Gamma_{0}$;
the total luminosity of the jet $L_{\rm j}=10^{53}\,{\rm erg\,s^{-1}}$
and the luminosity distance $d_{L} =4.85\times 10^{28}\,{\rm cm}\,(z = 2.0)$.
In order to intuitively compare the evolution of the photon energy and the electron energy with the radius, we assume the variation of the electron energy density with the radius as $\epsilon'_{\rm e}\propto r^{-2}$.
In addition, we adopt an on-axis observer in this paper.
Based on these settings, we perform numerical simulations and present their results in Figure~\ref{MyFig1}, which present the evolution of photon energy as a function of radius for the coasting ($\rm\delta = 0$), accelerating ($\rm\delta = 1/3$) and decelerating ($\rm\delta = -1/3$) jet cases, respectively. 
In the figure, the photon energy $\epsilon_{\rm r}$ and the radius $r$ are plotted.
The black dotted lines, which appear as solid lines due to the extremely dense distribution of data points, represent the results of MCRaT.
The black dash-dot-dot lines represent the analytical estimated photospheric radius.
The black dashed lines denote analytical estimates derived from the theoretical scaling relation $\epsilon_{\rm r} \propto r^{[-(\frac{2+\delta}{3})+\delta]}$ below the photospheric radius, while the temperature remains constant above the photospheric radius.

Figure~\ref{MyFig1} illustrates that the evolution of the photon energy with radius unfolds through three distinct, dynamically interlinked phases: the initial stage, the Compton scattering stage, and the steady stage. 
Initial stage:
Near the initial radius, the high optical depth facilitates frequent Compton scattering between photons and electrons. 
Due to the initial electron energy being set higher than the photon energy, the photon energy rapidly increases, as depicted by the steep rise of the black dotted lines.
Compton scattering stage:
After photons are heated by electrons to the peak energy, the photon energy begins to cool. 
At this stage, since the cooling rate of the electrons is set higher than the Compton scattering cooling rate of the photons, Compton scattering cooling dominates the photon cooling process, causing the photon energy to evolve as $\epsilon_{\rm r} \propto r^{[-(\frac{2+\delta}{3})+\delta]}$.
This behavior is evident in the region where the black dashed and dotted lines overlap. 
As the optical depth approaches unity (i.e., the photospheric radius $r_{\rm ph}$), the probability of Compton scattering diminishes, resulting in a declining photon cooling rate. 
The energy evolution of the photon  gradually begins to deviate from the relation 
$\epsilon_{\rm r} \propto r^{[-(\frac{2+\delta}{3})+\delta]}$, as shown in Figure~\ref{MyFig1}.
Steady-State stage:
With further jet expansion, the continued reduction in optical depth suppresses Compton scattering events, significantly slowing photon cooling. 
Ultimately, the photon energy reaches a steady state, remaining approximately constant with radius (as shown by the black dotted line), indicating a dynamic equilibrium between heating and cooling mechanisms.
It should be noted that the analytical scaling $\epsilon_r \propto r^{-(2+\delta)/3+\delta}$ is derived under the assumption $\Delta r_i/r_i \ll 1$, i.e., deep in the optically thick regime ($\tau \gg 1$).

In addition, we find that in the jet acceleration scenario, the photon energy reaches a steady state the fastest (i.e., the photon energy remains constant), while in the decelerating jet scenario, it reaches a steady state the slowest (see the subgraphs in Figure~\ref{MyFig1}). 
This indicates that the Compton scattering effect of photons is most significant during decelerating jets, and least significant during jet acceleration.

\section{Radiation spectrum of Photospheric emission in an accelerating/a decelerating jet.}\label{sec3}
In Section~\ref{sec2}, we present analytical estimates of the photon energy evolution with radius in accelerating and decelerating relativistic jets. 
Building on these results, this section has calculated the radiation spectrum of the photospheric model employing the same methodology as in \cite{2014ApJ...785..112D, 2020ApJ...899..111W}.
In our analytical estimates, we assume that the photon energy is proportional to the temperature.
The photon temperature is $T = D T'$, 
where $D = 1/[\Gamma(1 - \beta \cos\theta)]$ is the Doppler factor and $\theta$ is the angle between the radiation point and the line of sight.
The comoving temperature $T'$ is given by
\begin{equation}
T'(r) =
\begin{cases}
T_0 \left( \frac{r}{r_0} \right)^{-1}, & r_0 < r < r_{\mathrm{s}}; \\
T_0 \left( \frac{r_{\mathrm{s}}}{r_0} \right)^{-1} \left( \frac{r}{r_{\mathrm{s}}} \right)^{-(2+\delta)/3}, & r_{\mathrm{s}} < r < r_{\mathrm{ph}}; \\
T_0 \left( \frac{r_{\mathrm{s}}}{r_0} \right)^{-1} \left( \frac{r_{\mathrm{ph}}}{r_{\mathrm{s}}} \right)^{-(2+\delta)/3} , &   r_{\mathrm{s}} < r_\mathrm{ph} < r .
\end{cases}
\label{eq:temperature_evolution}
\end{equation}
Here, $T_0 = \left[ L_{\rm j} / 4\pi r_0^2 a c \right]^{1/4}$ is the base jet temperature at $r_0$, and $a$ is the radiation density constant.
$r_{\mathrm{ph}}$ is the photosphere radius, defined as the location where the optical depth $\tau = 1$, with $\tau$ calculated using the method described in \cite{2020ApJ...899..111W}.
The jet parameters are the same as those used in the numerical simulations described above. 
In these simulations, we did not employ the hydrodynamic module of the code; instead, the jet parameters were prescribed directly. Additionally, we present the radiation energy spectrum obtained using the MCRaT.

Figure ~\ref{MyFig2} present the normalized radiation spectra
($\epsilon_{\rm r}F/\epsilon_{\rm p}F_{\rm max})-(\epsilon_{\rm r}/\epsilon_{\rm p})$ obtained based on Equation~(\ref{EQ:photon energy K}) and the MCRaT simulation, respectively.
Here, $\epsilon_{\rm r}$ denotes the observed photon energy, $\epsilon_{\rm r}F$ is the energy flux,  $\epsilon_{\rm p}F_{\rm max}$ represents the peak flux, and $\epsilon_{\rm p}$ is the photon energy corresponding to the peak flux.
We examine three cases: coasting ($\rm\delta = 0$), accelerating ($\rm\delta = 1/3$), and decelerating ($\rm\delta = -1/3$) jets, represented by black, red and blue lines in Figure~\ref{MyFig2}, respectively. 
The solid lines show the analytical predictions, while the dotted lines with error bars indicate the simulation data of MCRaT; the error bars are estimated using Poisson statistics to mimic observational uncertainties.
Figure~\ref{MyFig2} shows that the spectral width depends on the jet acceleration index $\delta$. 
Specifically, the spectrum for the decelerating case ($\delta=-1/3$, blue line) is broader than that of the coasting case ($\delta=0$, black line). 
Similarly, the spectrum corresponding to $\delta=0$ is wider than that of $\delta= 1/3$ (the red line in Figure ~\ref{MyFig2}).
This indicates that a smaller $\delta$ results in a broader radiation spectrum.

The broadening of the energy spectrum mainly occurs because, when the jet is decelerated, the beaming effect is weakened. 
This allows high($> \epsilon_{\rm p}$)- and low-energy($< \epsilon_{\rm p}$) radiation from high latitudes (larger $\theta$) to become visible to the observer.
There is another factor causing spectral broadening: if the photon energy cools too rapidly with radius, the contribution from upstream radiation near the photospheric radius increases, leading to broadening in the high-energy portion of the spectrum. 
For instance, in Figure ~\ref{MyFig2}, the discrepancy between the slope of the blue solid line and the corresponding data points from numerical simulations is larger compared to the cases of coasting (black solid line) and acceleration (red solid line). 
Consequently, the analytically calculated spectrum ( solid blue line in Figure ~\ref{MyFig2}) in the high-energy region reflects the superposition of two broadening effects, resulting in a flatter spectrum in the high-energy band.

In addition, Figure~\ref{MyFig2} shows that the analytically estimated spectrum agrees well with the numerical simulation around the peak frequency, while deviations are observed in the high- and low-energy bands.
This discrepancy is mainly attributed to the slight difference in the radial evolution of the photon energy between the analytical estimation and the numerical simulation.
In analytical estimation, the photon energy decreases sharply at the radius where the optical depth equals unity ($\tau = 1$) and subsequently remains constant, 
as described by Equation~(\ref{eq:temperature_evolution}).
In contrast, the numerical simulation exhibits a gradual decrease in photon energy, 
as illustrated by the black dotted line in Figure~\ref{MyFig1}. 
This behavior results in a flatter radiation spectrum across both the high- and low-energy bands.


\section{Summary and Discussions}\label{discussion and summary}\label{sec4}
The cooling of photons due to Compton scattering near the photospheric radius is important for understanding photospheric emission.
Previous studies show that photon cooling in a coasting jet is affected by Compton scattering, due to the slight misalignment of the velocity vectors of the scattering electrons.
Recognizing that jets often undergo acceleration or deceleration, we extend previous studies by examining Compton scattering effects in non-stationary jet environments.
Assuming a power-law evolution of the Lorentz factor ($\Gamma\propto r^{\delta}$), we develop an analytical framework to estimate the photon energy evolution below the photospheric radius (i.e., where the optical depth is larger than unity): $\epsilon_{\rm r} \propto r^{[-(\frac{2+\delta}{3})+\delta]}$.
Comprehensive numerical simulations performed with the MCRaT confirmed the validity of our analytical approach in various jet dynamics, particularly for cases with different values of $\delta = $ 1/3 and -1/3, which represent different jet acceleration/deceleration regimes.
We find that, below the photospheric radius, the numerical simulation results are consistent with those obtained from the analytical estimation.

On the basis of the analytical estimation of the photon energy-radius relationship, we derive the radiation spectrum of photospheric emission through analytical calculations. 
We find that a smaller $\delta$ leads to a broader radiation spectrum.
The broadening of the energy spectrum primarily occurs as the jet decelerates, weakening the beaming effect, which reduces the dominance of high-energy radiation and allows lower-energy emissions to contribute more significantly to the observed spectrum. 
Additionally, rapid cooling of photon energy with an increasing radius may cause spectral broadening in the high-energy band.

\begin{figure*}[ht]
     \centering
    \begin{minipage}[b]{0.55\textwidth}
        \centering
        \includegraphics[width=\textwidth]{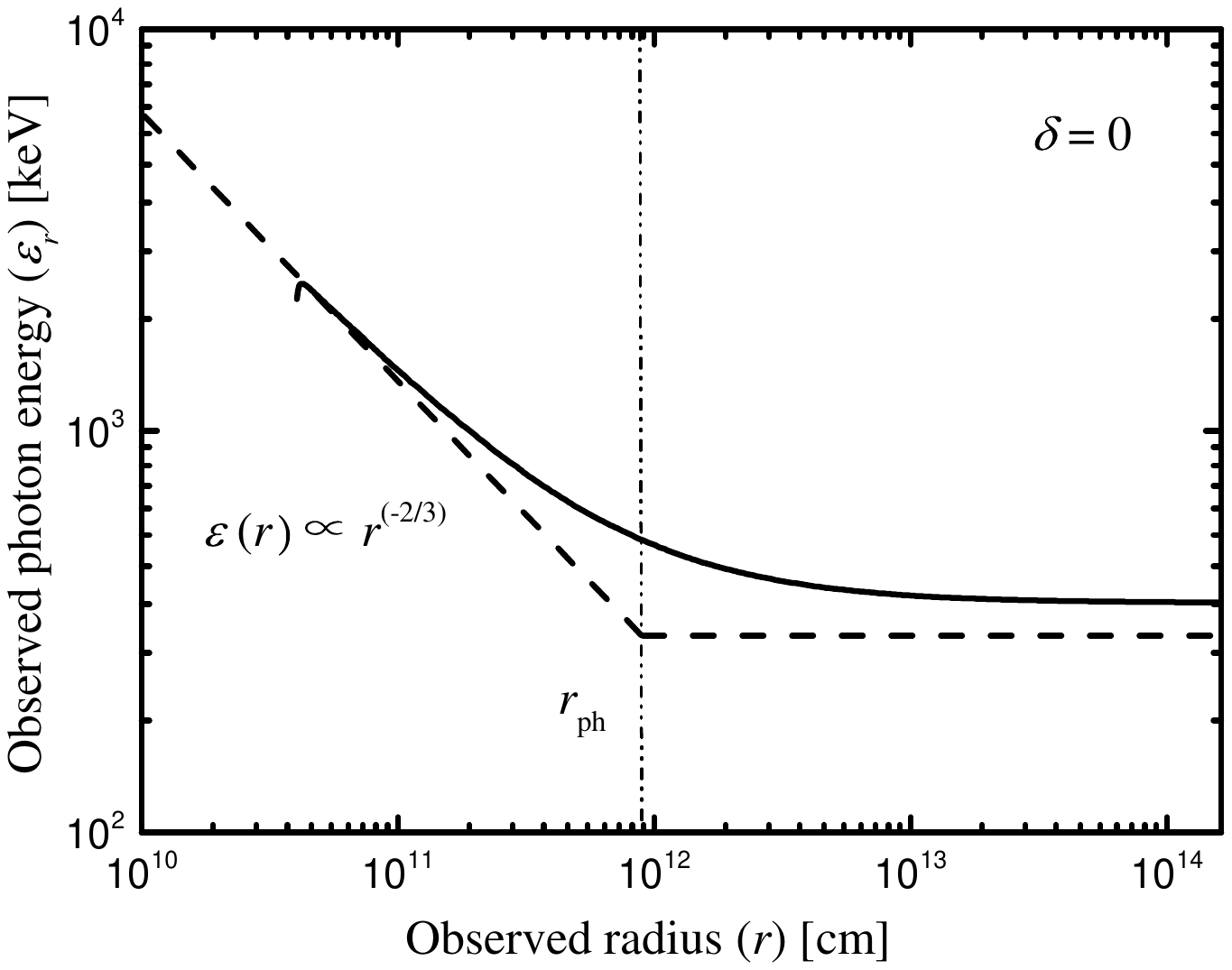}
    \end{minipage}
    \hfill
    \begin{minipage}[b]{0.55\textwidth}
        \centering
        \includegraphics[width=\textwidth]{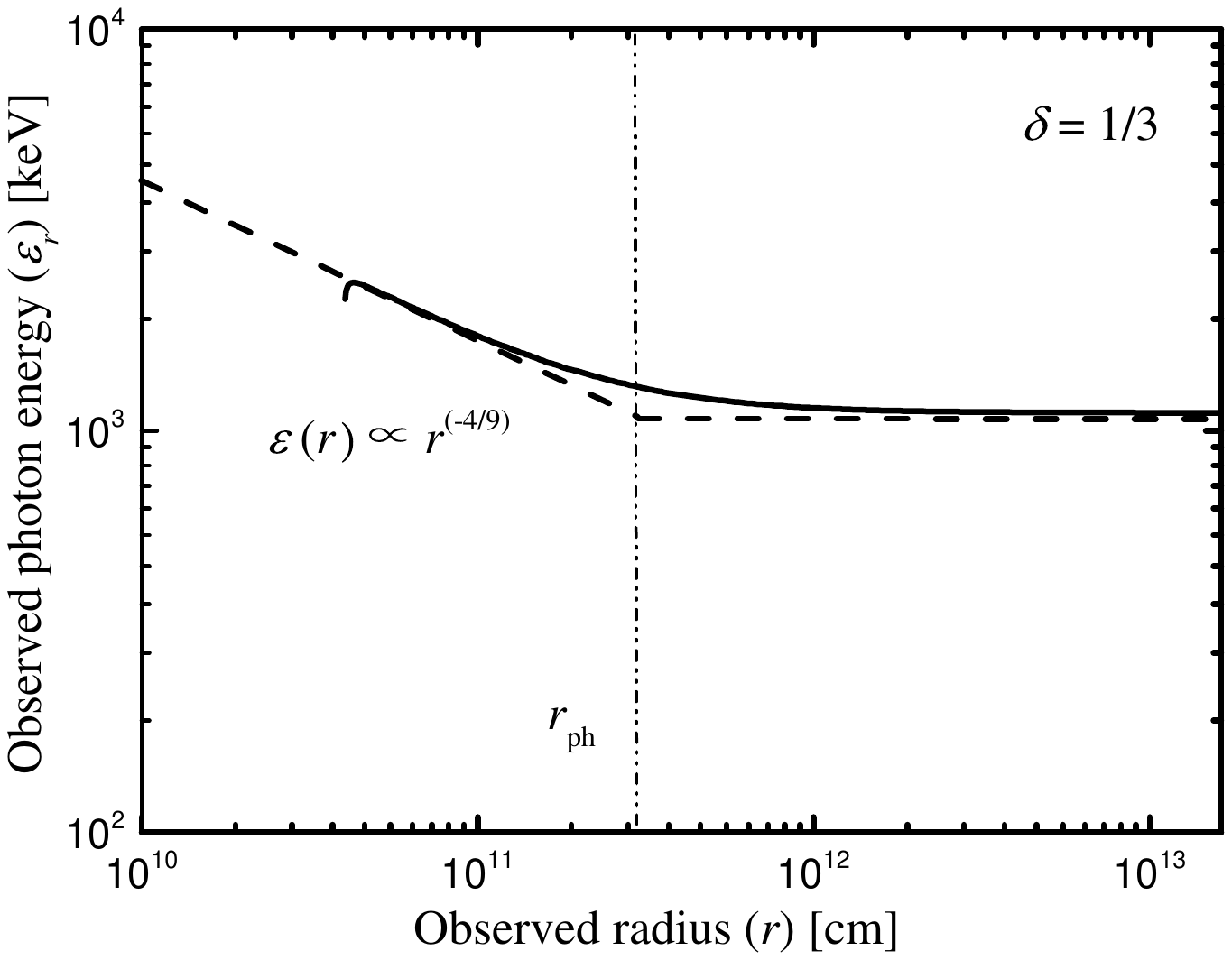}
    \end{minipage}
    \hfill
    \begin{minipage}[b]{0.55\textwidth}
        \centering
        \includegraphics[width=\textwidth]{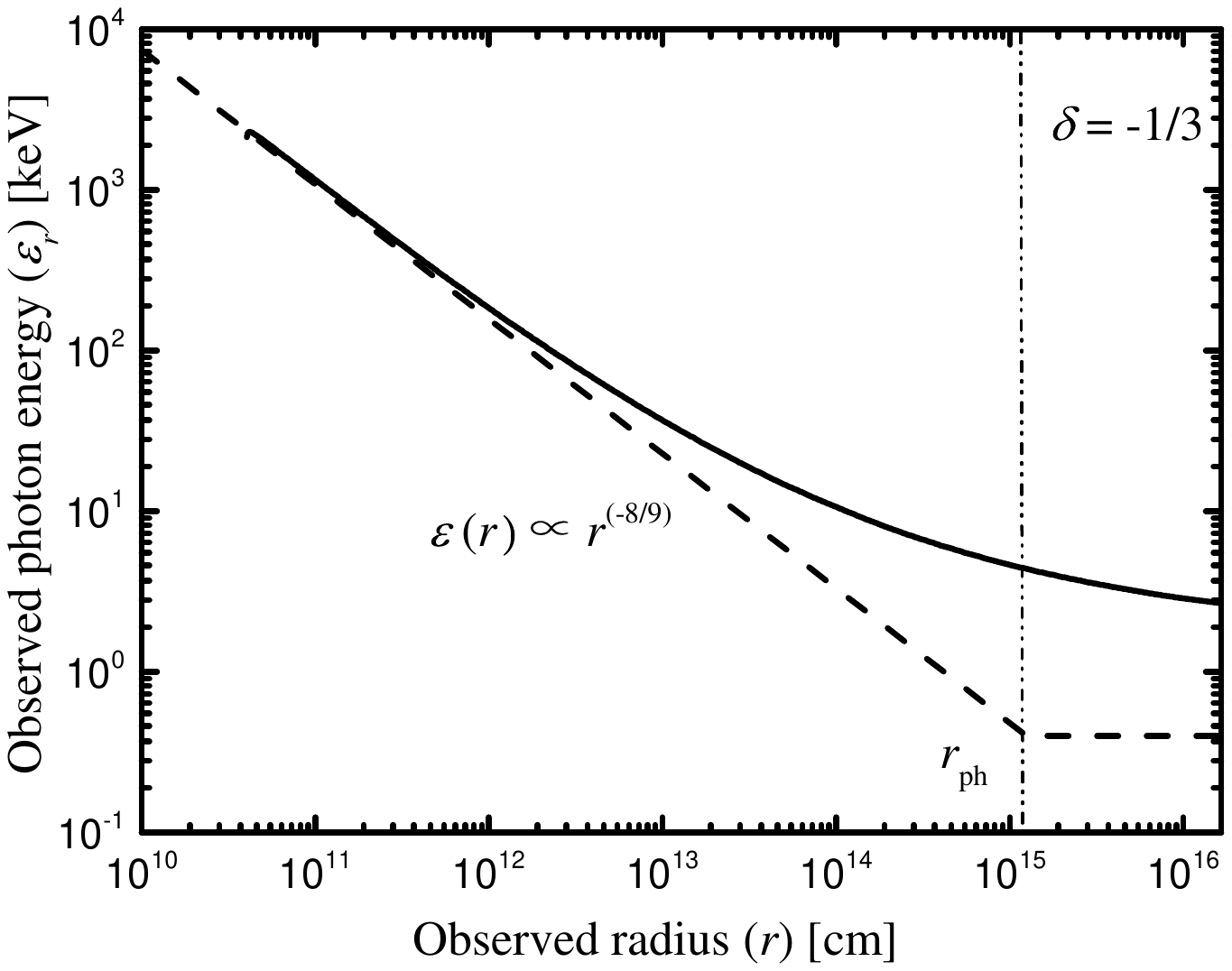}
    \end{minipage}
    \caption{Relationship between photon energy $\epsilon_{\rm r}$ and radius a
coasting ($\delta=0$, the upper panel), an accelerating ($\delta=1/3$, the middle panel) and a decelerating ($\delta=-1/3$, the lower panel) jets scenarios, respectively, obtained from simulations of MCRaT.
The horizontal and vertical axes represent the normalized radius $r$ and photon energy $\epsilon_{\rm r}$ in the observer frame, respectively.
The black dotted lines show the numerical simulation results.
The black dash-dot-dot lines represent the analytical estimated photospheric radius.
The black dashed lines denote analytical estimates derived from the theoretical scaling relation $\epsilon_{\rm r} \propto r^{[-(\frac{2+\delta}{3})+\delta]}$ below the photospheric radius, while the temperature remains constant above the photospheric radius.}
    \label{MyFig1}
\end{figure*}

\begin{figure}[ht]
\centering
\includegraphics[width=10.5 cm]{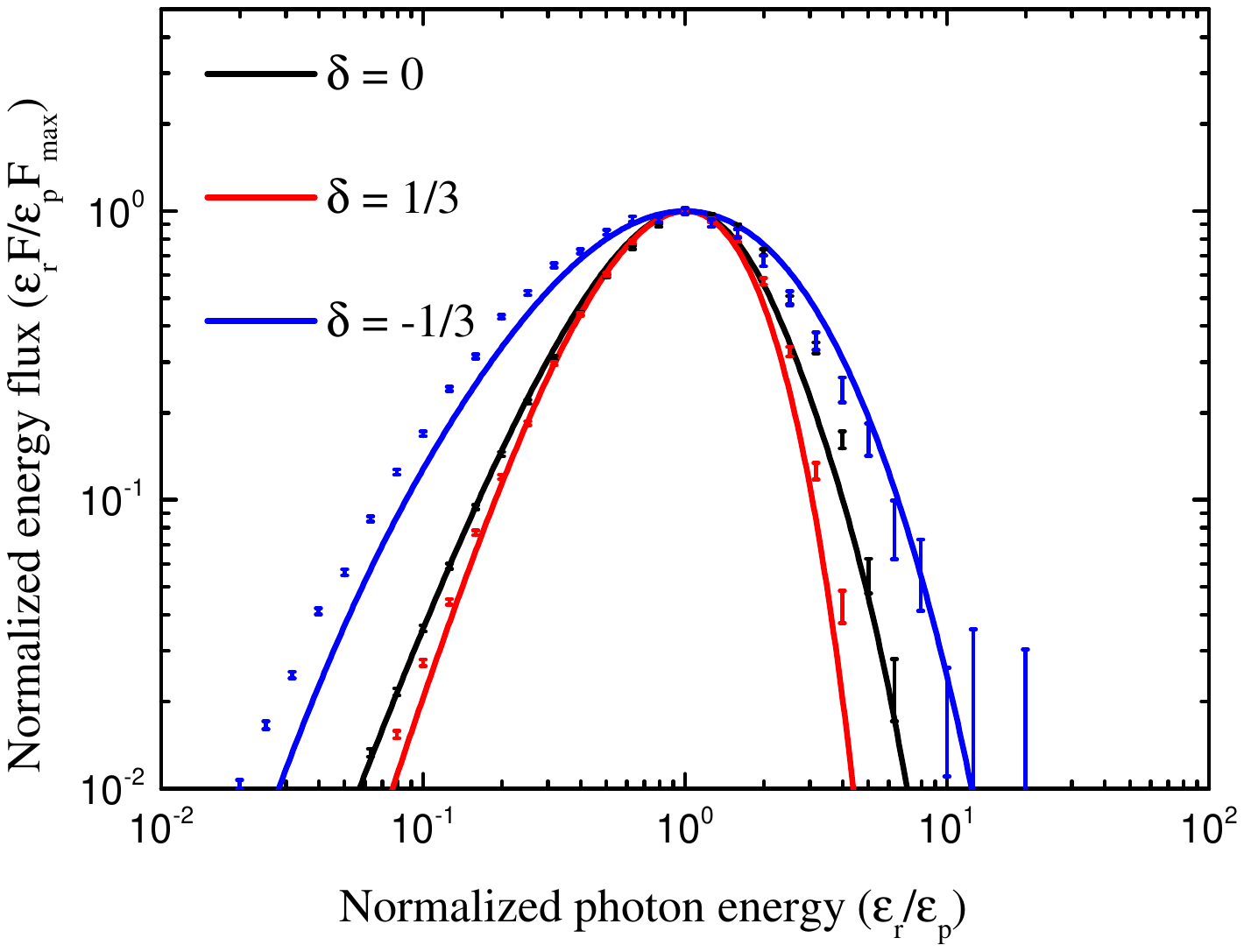}
\caption{The radiation spectra obtained from analytical estimation based on 
equation ~(\ref{EQ:photon energy K}) and MCRaT's simulations for
coasting ($\delta=0$), an accelerating ($\delta=1/3$) and a decelerating ($\delta=-1/3$) jet cases, respectively.
The horizontal and vertical axes are the normalized radiation energy bands ($\epsilon_{\rm r}/\epsilon_{\rm p}$) and energy flux ($\epsilon_{\rm r}F/\epsilon_{\rm p}F_{\rm max}$), respectively, 
where $F_{\rm max}$ is the peak flux and $\epsilon_{\rm p}$ is the photon energy corresponding to the peak flux.
The solid lines are the result of analytical estimation, and the dotted lines with error bars are the results of numerical simulations by the MCRaT.
The black solid line and dotted line are the radiation spectra when the index $\delta=0$;
the red solid line and dotted line represent the energy spectrum when the jet is accelerating ($\delta=1/3$); when the jet decelerated ($\delta=-1/3$), the radiation spectrum are represented by the blue solid line and dotted line.
\label{MyFig2}}
\end{figure}

\begin{figure}[ht]
\centering
\includegraphics[width=9.5 cm]{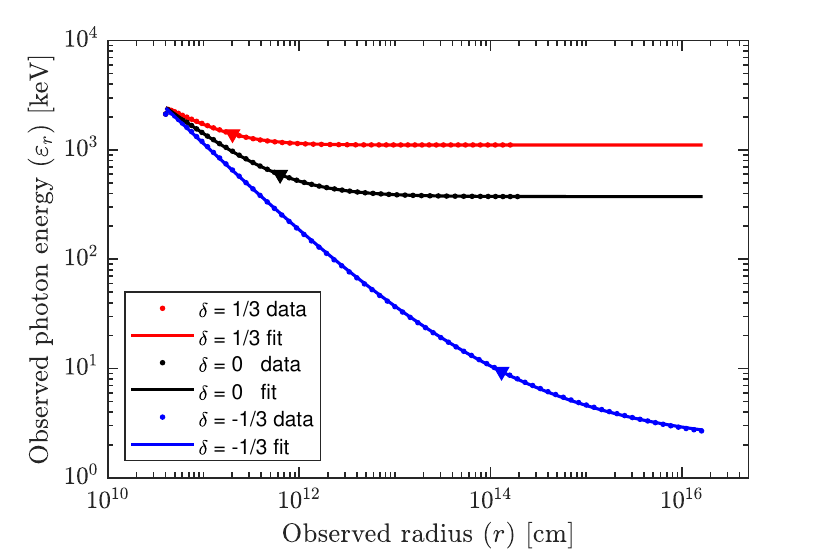}
\caption{The radial evolution of the observed photon energy $\varepsilon_r$ as a function of the observed radius $r$. Dotted curves represent the numerical simulation data, while solid lines show the fits using a smooth broken power-law function $\epsilon_r(r) = A r^{\alpha} \left[ 1 + \left( \frac{r}{r_c} \right)^{\beta} \right]^{-\alpha/\beta}$. 
Results are presented for three different values of the parameter $\delta$: $1/3$ (red), $0$ (black), and $-1/3$ (blue). 
Here, the triangle symbols mark the positions corresponding to the characteristic transition radius $r_c$.}
\label{MyFig3}
\end{figure}

In addition, we present the radiation spectrum obtained from numerical simulations using the MCRaT.
Figure~\ref{MyFig2} shows that the analytically estimated spectrum agrees well with the numerical simulation around the peak frequency, while deviations are observed in the high- and low-energy bands.
This discrepancy is mainly attributed to the slight difference in the radial evolution of the photon energy between the analytical estimation and the numerical simulation.
As the optical depth approaches unity ($\tau \sim 1$) near the photospheric radius, the mean free path increases and the condition $\Delta r_i/r_i \ll 1$ breaks down, causing the photon energy evolution to gradually deviate from the power-law prediction. 
This transition is clearly visible in Figure~\ref{MyFig1}, where the numerical data (dotted curves) start to flatten before reaching the analytical photospheric radius.
To quantitatively characterize the full radial profile, we fit the numerical data with a smooth broken power-law function of the form
$\epsilon_r(r) = A r^{\alpha} \left[ 1 + \left( \frac{r}{r_c} \right)^{\beta} \right]^{-\alpha/\beta}$,
where $\alpha = -(2+\delta)/3 + \delta$ is fixed to the theoretical exponent, $r_c$ is the characteristic transition radius (close to the photospheric radius), $\beta$ controls the sharpness of the transition, and $A$ is a normalization constant. The best-fit curves are shown as colored solid lines in Figure~\ref{MyFig3}. 
The fitted parameters for the cases $\delta = 1/3$, $0$, and $-1/3$ are as follows: $A = 2.1\times10^{8}\ \mathrm{cm}$, $2.8\times10^{10}\ \mathrm{cm}$, and $3.4\times10^{12}\ \mathrm{cm}$, respectively; $r_c = 2.0\times10^{11}\ \mathrm{cm}$ ($\tau\sim1.0$), $6.32\times10^{11}\ \mathrm{cm}$ ($\tau\sim1.5$), and $1.31\times10^{14}\ \mathrm{cm}$ ($\tau\sim3.0$), respectively; $\beta = 1.4$, $1.0$, and $0.4$, respectively.
Although the fitting results confirm that the power-law segment holds for $r \ll r_c$, while a smooth rollover to a constant plateau occurs at $r \gtrsim r_c$, and the fitted $r_c$ values are close to the analytically estimated position where the optical depth is around unity near the photosphere---indicating that the transition is governed by the same physical mechanism---there remains a discrepancy between the fitted $r_c$ and the photospheric location. Moreover, the physical meaning of the parameter $\beta$ is not clear. Therefore, further in-depth investigation is needed if one intends to describe the radial evolution of the photon energy using a smooth broken power-law function.

\clearpage

\acknowledgments{The authors of this research contributed the following: 
conceptualization, Hao-Qiang Zhang and Da-Bin Lin; methodology, Hao-Qiang Zhang; software, Guo-Yu Li; investigation, Hao-Qiang Zhang; writing---original draft preparation, Hao-Qiang Zhang; supervision, Da-Bin Lin; funding acquisition, Da-Bin Lin. 
This work is supported by the National Natural Science Foundation of China (grant Nos. 12494575 and 12273005), the special funding for Guangxi Bagui Youth Scholars (Da-Bin Lin), the Guangxi Talent Program (“Highland of Innovation Talents”), the Guangxi Science and Technology Innovation Platform Program (Leitai Action Plan, Grant No. Guike LT2600640026 ), and Guangxi Key R\&D Program (Guangxi Funeng Action Plan, Grant No. FN2504240030).}

\clearpage
\bibliographystyle{apj}
\bibliography{bibliography}
\end{document}